\documentclass[10pt]{article}
\usepackage[latin1]{inputenc}
\usepackage{amsmath}
\usepackage{amsfonts}
\usepackage{amssymb}
\usepackage{mathrsfs}
\usepackage{subcaption}
\usepackage{booktabs}
\usepackage{authblk}

\usepackage{fancyhdr}

\usepackage{blindtext}

\usepackage{float}
\usepackage{fancybox,graphicx}
\usepackage{caption}
\usepackage{color}
\usepackage{easyReview}

\usepackage[colorlinks]{hyperref}

\usepackage{accents}
\usepackage[titletoc,title]{appendix}
\usepackage{cite}

\usepackage{mathtools}

\usepackage[top=2in, bottom=1.5in, left=1in, right=1in]{geometry}

\title{Weighted complement graphs of spatial networks with functional connections reveal nodes with high potential for new links}

\author[1]{Tina \v Sfiligoj$^*$} 

\affil[1]{Faculty of Maritime Studies and Transport, University of Ljubljana, Slovenia, e-mail: tina.sfiligoj@fpp.uni-lj.si}

\author[2]{Oded Cats}

\affil[2]{Department of Transport \& Planning, Delft University of Technology, The Netherlands}

\begin{document}

    \maketitle

\section*{Abstract}

In this study, we take a systematic look at the unrealised part of public transport networks (PTNs) with functional connections. We consider their complement graphs and study their structure. The complement graph $\overline G$ of an unweighted graph $G$ is a straightforward concept, yielding a graph on the same set of nodes, and an edge exists in $\overline G$ if and only if it is not present in $G$. In contrast, a weighted complement graph cannot be uniquely determined. However, if we consider PTNs with travel times as edge weights, there are physical constraints on the possible weight ranges. We propose a method to construct weighted complement graphs of operational PTN graph representations based on the geographical distances between nodes (representing stops) and assign weights to edges based on distance, combined with network-specific distributions of effective velocities and waiting times. We observe that the most central nodes in the weighted complement graph do not correspond to the least central nodes in the original network but are, remarkably, those in the geographical centre of the network that lack topological connectedness. Testing against null models on a dataset of 31 metro networks worldwide confirms that this is a fundamentally spatial effect. 


\textbf{Keywords:} public transport, network science, node centrality, complement graph, spatial network

\section{Introduction}

The analysis of node importance in spatial networks \cite{barthelemy2011spatial}, and in network analysis in general, often focuses on identifying the most important nodes in the network, which correspond to critical locations in the network, be it stops in public transport networks or junctions in road networks, among others. The node importance is measured by any of the numerous centrality measures, among which the most common are degree, closeness, betweenness and eigenvector centrality. The highest centrality nodes are those with the greatest influence on various important phenomena on networks, such as robustness or vulnerability (e.g., \cite{liu2017recognition, freitas2022graph, palk2026robustness}), or identifying the most influential spreaders (e.g., \cite{liu2013ranking, pei2013spreading, kandhway2016using}).

In real-world spatial networks, such as transportation networks, the topology itself is often not a sufficient descriptor of the system due to its geographical embeddedness and the geometrical properties of the geographical space. The inclusion of infrastructural or operational properties, which are added as edge weights (e.g., travel times or service frequency), has consistently been shown to more accurately reflect the actual behaviour of the system, including the assessment of node importance (e.g., \cite{luo2020can, vsfiligoj2025node}. Including edge weights in the models introduces non-trivial relations between node importance in unweighted networks and their weighted counterparts (e.g., node degree vs. strength, or betweenness centrality with travel cost on the links vs. unweighted edges) (e.g., \cite{opsahl2010node, oldham2019consistency}). This is especially prominent in spatial networks where the edge weights, representing travel costs (distance or travel time), are constrained based on the physical embeddedness of the network. These relationships are often non-linear, indicating a complex interplay between topology and space \cite{barrat2005effects, barthelemy2011spatial}.


Understandably, past work has heavily focused on investigating the most central and thereby important nodes. Less central nodes could have their lack of centrality attributed to either spatial features (e.g., peripheral location) or network features (e.g., poor connectivity despite a geographically favourable location). We argue here that the distinction between these two categories is important and consequential for network development. Notably, the nodes in the latter category may be currently unimportant for the prevailing network structure but not unimportant when considering potential future network states.

In the area of transport networks, \cite{batac2022shortest} and \cite{cirunay2023evolution} considered road networks and identified the least important nodes according to betweenness centrality, which they called the topological periphery. They found that the least important nodes fall into two categories: those at the geographical periphery and poorly connected nodes in the geographical centres (often representing cul-de-sacs). Moreover, the shortest paths connecting the least topologically important but geographically central nodes are those with the highest circuity, indicating significant inefficiencies in the network. While this work differs from our approach in considering road networks with their inherent restrictions to planarity, it offers an interesting result which we comment on in the Discussion section.

Public transport networks (PTN) are a prominent class of spatial networks, with stops embedded in a geographical space that imposes physical constraints on system operation. Even in the theoretical ideal case without resource constraints, the geographical distance between a pair of nodes poses an upper limit on the minimum travel time between them, as the maximum velocity and acceleration along a link are bounded by technology, traffic and passenger safety concerns. This embeddedness, together with the actual PTN infrastructure and service frequencies, sets the feasible ranges of travel times, which are often included as edge weights in PTN modelling.

Although there is a large body of literature on the interplay between topology and geography in spatial networks, related work mostly considers planar networks that represent network infrastructure. In PTNs, this is the so-called L-space \cite{von2009public}, or space-of-infrastructure representation \cite{luo2020can}, in which two node share an edge if they represent consecutive stops on a line. 
In comparison, spatial networks with functional connections are less common in the literature. By functional connections, we mean any relation between nodes stemming either from operational features of the network, such as line operations, or dynamics on networks, such as passenger flows. Such edges carry a level of spatial abstraction, in that they give rise to highly non-planar structures. In PTN analysis, a widely-used representation is the so-called P-space representation \cite{von2009public}, or space-of-service \cite{luo2020can}, where nodes represent stops, and there is an edge between two nodes if the corresponding stops lie on the same line. In other words, each line is a clique, and paths in this representation effectively count the number of transfers. Edge weights in P-space often represent service frequencies, which can be translated into average waiting times. In addition, from the underlying infrastructure, in-vehicle times along P-space edges can be computed from the corresponding paths in the infrastructural representation; thus, we can obtain total travel times between all pairs of nodes.

Such non-planar networks can have high edge densities, and in the P-space, any - existing or potential - edge is understood in terms of the two stops sharing a line. In this representation, the absence of an edge does not imply a physical impossibility but rather an unrealised direct connection. This leads us to ask: can we gain original insights into the network properties by systematically analysing the unrealised part of the network? i.e. all the services/connections that could have existed but are not included in the current PTN.
In doing so, we encountered and addressed an additional research question: how can we use the spatial embeddedness of the network to estimate possible weighted realisations of currently non-existing edges?

We approach the problem by considering complement graphs of the functional representations of spatial networks. Given a graph $G(V,E)$ with a node set $V$ and an edge set $E$, and the set of all possible edges $E^U$, $|E^U|=\frac{n(n-1)}{2}$, its \textit{complement graph} is defined as $\overline G(V, \overline E)$, where $\overline E = E^U - E$; i.e., the graph on the same set of nodes, and there is an edge between two nodes in $\overline G$ if and only if it is not present in $G$. The complement graph of an unweighted graph is a straightforward concept. In contrast, weighted complement graphs cannot be uniquely determined. However, with regard to the physical constraints of the spatial networks, there exist ranges of feasible values that allow us to estimate the edge weights in the complement graph.

In this study, we propose a method to construct weighted complement graphs of the P-space representation of PTNs based on the geographical distances between nodes (stops), and assign edge weights - representing total travel times - based on the distances and network-specific distributions of effective velocities and waiting times. We then compute and compare centrality measures in the original P-space graph $G$ and its complement $\overline G$. In $G$, the nodes with the highest centrality are the most well-connected nodes in the network and represent transport hubs. Specifically, we consider degree and eigenvector centrality, and for their weighted calculation transform the cost-type travel time weights $c_{ij}$ to strength-type weights using the inverse transformation $s_{ij} = 1/c_{ij}$. In $\overline G$, we interpret the nodes with the highest centrality as those with the highest potential for new edges. Since connections in the complement graph represent, by definition, exactly the connections absent in the original graph, the most central nodes  -- i.e., those that have the most connections -- are the ones that have the most missing edges in the original graph. In the unweighted case, the degree in $\overline G$ is a linear decreasing function of the degree in $G$: if a node $i$ has degree $k$ in $G$, its degree will be $n-k-1$ in $\overline G$, which is exactly the potential for new links. In the case of eigenvector centrality, the measure also includes the importance of node's neighbours; i.e., the potential for connections between nodes that are poorly connected in $G$. We furthermore argue that including the spatial locations of the nodes provides a more reliable identification of nodes with high potential for new connections based on their geographical proximity.

To the best of our knowlegde, the systematic analysis of a complement graph as an object in its own right, especially its weighted version, has not been considered before in the context of public transport networks. Moreover, the literature in other areas of network science applications is scarce. An important work in this direction is \cite{everett2014networks}, where the authors considered complement graphs of social networks, interpreting the edges in the complement graph as negative ties, i.e., adversarial relationships between individuals. They use the complement graph to study the cohesiveness of the network structure with respect to those negative ties. However, they construct an unweighted network; indeed, in social networks, the hard physical constraints on possible edge weights are absent. A version of the weighted complement network was however proposed with edge weights based on the importance of incident nodes.

We note that there is a substantive amount of work considering unrealised edges in transport planning problems, such as transport network planning problem (\cite{chen2011transport, schobel2012line, cats202550}), and methods that consider missing edges, such as link prediction in graph neural networks (GNN) (e.g., \cite{shanthappa2024origin, mouronte2021modeling}). While these approaches carry important implications for PTN planning, they function at the level of individual edges, thereby inhibiting the potential for a structural analysis of the unrealised parts of the network.


Our study complements the existing literature by considering a different approach based on analysing the complement graphs of P-space PTNs.
The construction of the weighted complement network builds on a concept similar to the idea of the idealised network on the same set of geographically positioned nodes. The notion of the idealised network was introduced in \cite{latora2001efficient} for a reliable measure of efficiency and was recently used in the formulation of PTN accessibility in \cite{vsfiligoj2026network}. The idealised network is a complete graph; i.e., there is an edge between each pair of nodes. The travel time weights are then estimated based on the geographical distance between nodes and the expected or feasible values of speed on links and waiting times. While the idealised infrastructural network is infeasible to realise, in representations with functional connections, such as P-space, the absent edges are in principle all feasible; in our case, they are candidates for new lines in the network which utilise the existing infrastructure.







\medskip

\begin{figure}[H]
	\begin{center}
		\texttt{}\includegraphics[scale=0.4]{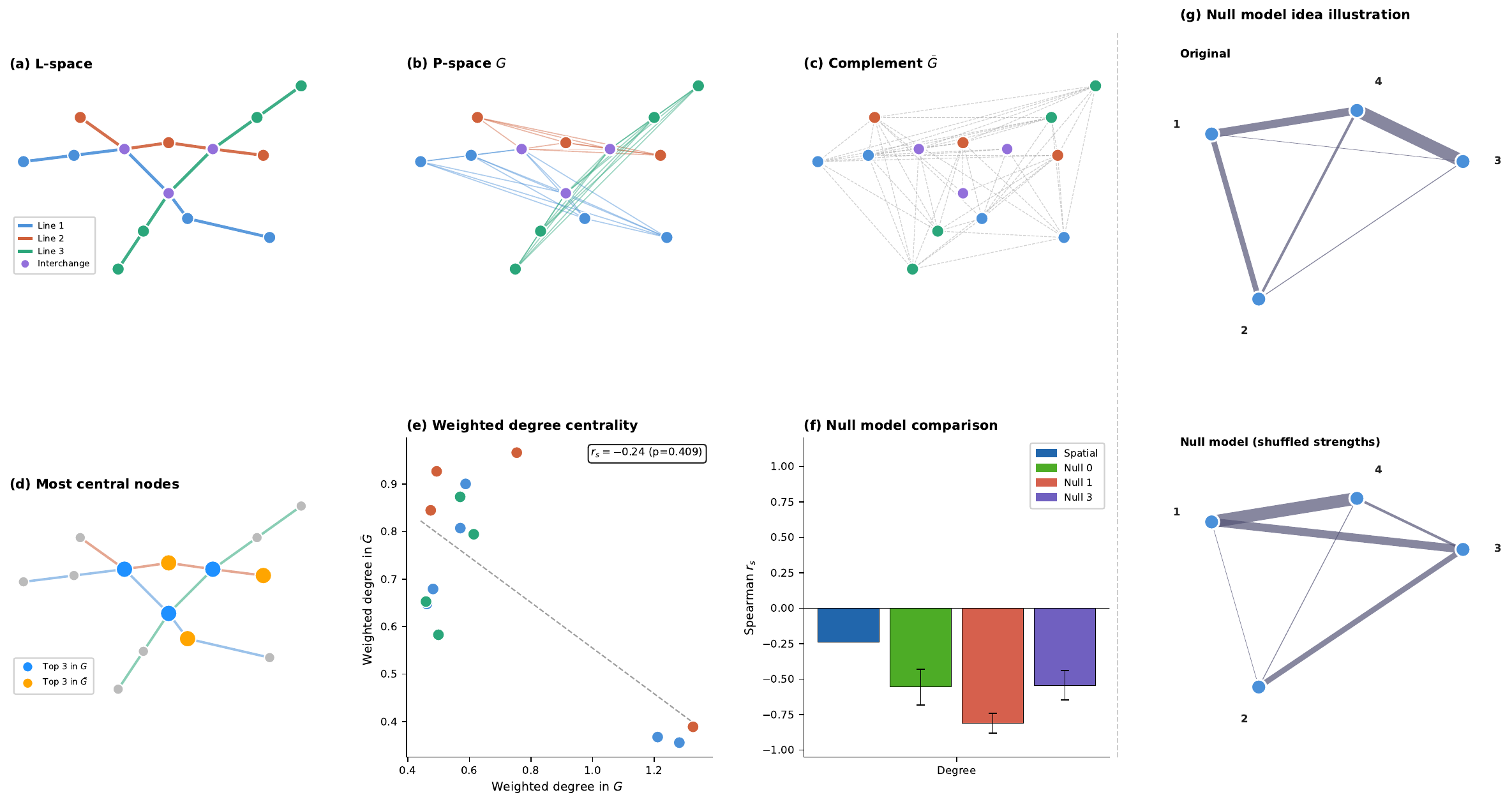}
	\end{center}
	\caption{The illustration of our methodology on a simple toy network with three lines embedded in geographical space. Panel $(a)$ shows the L-space representation with colour-coded lines, and panel $(b)$ shows the graph $G$ corresponding P-space representation, where each line is a clique. We work with the complement graph of $G$, $\overline G$, which is shown in panel $(c)$. The edge weights in both $G$ and $\overline G$ represent total travel times (i.e., in-vehicle time plus average waiting time); in $G$ travel times are obtained from data, and in $\overline G$ they are estimated based on the geographical distance between nodes, divided by network-specific effective velocity on links, plus network-specific average waiting time. In the next step, we calculate node centrality in $G$ and $\overline G$; in this case degree centrality, for which the inverse transformation of edge weights is performed. In panel $(d)$, the three most important nodes in $G$ and $\overline G$ are shown in blue and orange, respectively, and panel $(e)$ shows the scatter plot of node centrality in $\overline G$ vs. $G$ (in this plot, each point represents a stop the points are colour-coded according to the line-colouring scheme in panels $(a)$ and $(b)$. We observe that the relationship is higly non-linear, which is in remarkable contrast to the negative linear relationship in the unweighted case. In addition, we calculate the Spearman correlation coefficient ($r_S=-0.24$), and the correlation strength is not significant). A remarkable result is that the most important nodes in $\overline G$ are not those on the geographical periphery. We posit that this is the effect of the interplay between topology and geography. We test this by comparing to null models where the degree sequences are preserved and edge weights do not correlate to geographical distances between nodes. We compute the Spearman correlation coefficient for each model; the results are shown in panel $(f)$. The main idea behind the null models is shown on a separate small network in panel $(g)$. The original network (top) is a clique on four nodes, and the edge cost $c_{ij}$ equals the Euclidean distance between nodes. This is then converted to strength $s_{ij} = 1/c_{ij}$; i.e., nodes that are closer in space will have edges with higher strengths as visualised via edge thickness (thicker edge means higher strength). In the null model (bottom of panel $(g)$), the edge weights are reshuffled so that they no longer correlate to Euclidean distances as is seen for edges (1, 3) or (2, 3) which have high strengths despite the nodes being far apart. Moreover, the strength of edge (3, 4) is low, although these are the nearest nodes. In this way we disentangle topology from geography. In the analysis, we use additional null models where also the topology is changes by degree-preserving edge reshuffles.}
	\label{fig:toy}
\end{figure}

The contribution of this study is twofold. First, we develop a methodology for constructing weighted complement networks of spatial networks with functional connections. Second, we systematically analyse the interplay between spatial and topological structure and its effect on the relationships between node importance in the weighted network and its weighted complement counterpart. To disentangle topology from geography, we test the results against several null models, in which edge weights are not correlated to spatial distance between nodes. We apply the methodology to a dataset of 31 metro networks worldwide. Our approach is illustrated for a simple toy network in Figure \ref{fig:toy}.

\section{Results}

The methodology is illustrated for the Vienna metro network in subsection \ref{sec:results_vienna}. The detailed results for the complete dataset of 31 metro networks are given as supplementary materials. Subsection \ref{sec:null_all_cities} presents the results of testing against null hypotheses for the full dataset.

\subsection{Case study: the Vienna metro}
\label{sec:results_vienna}

\begin{figure}[H]
	\begin{center}
		\texttt{}\includegraphics[scale=0.65]{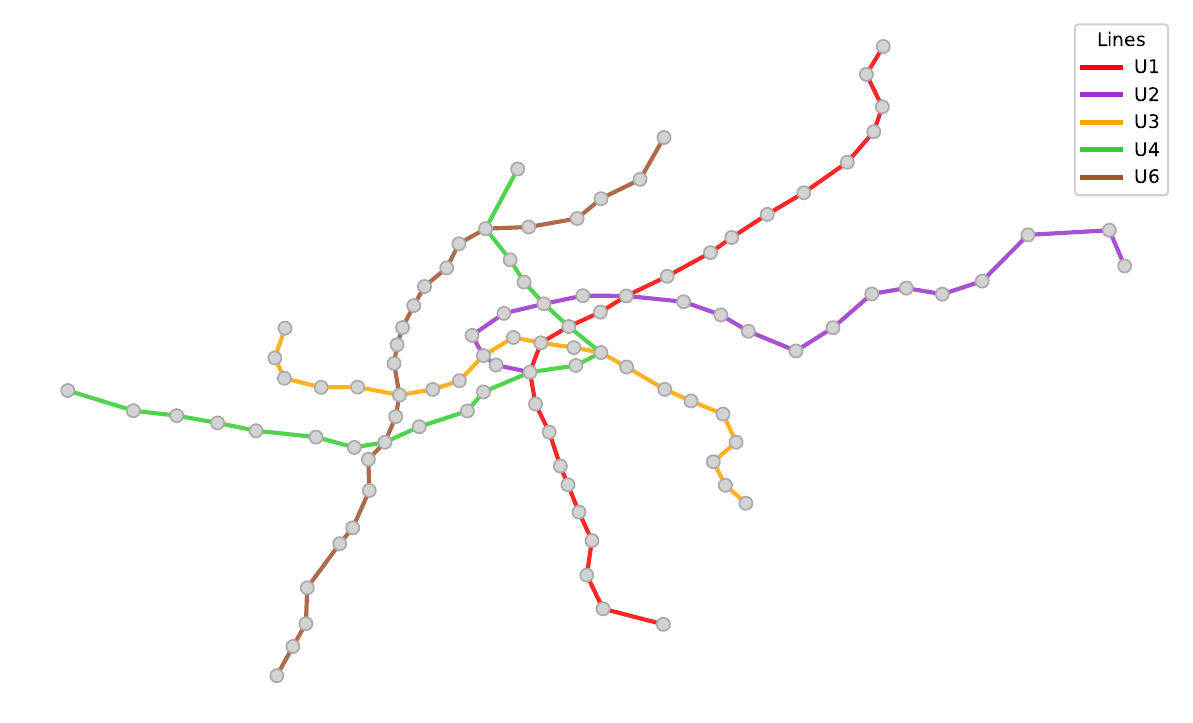}
	\end{center}
	\caption{Vienna L-space with colour-coded lines.}
	\label{fig:vienna_l_space}
\end{figure}

The Vienna metro network has $n=98$ stops (nodes) and operates five lines (U1 - U4 and U6, U standing for U-bahn). The L-space network with colour-coded lines is shown in Figure \ref{fig:vienna_l_space}.

The density of the P-space representation of the Vienna metro is $\approx24\%$, and consequently, the density of its complement is $\approx76\%$. While the complement density is still significantly higher than that of the original network, they are of the same order of magnitude, in stark contrast to the L-space representation, where the densities of the original and the complement graphs are $\approx2\%$ and $\approx98\%$, respectively.

The geographical span (i.e., the largest geographical distance between a pair of nodes) of the Vienna metro is $\approx 15$ km (the largest geographical distance between a pair of nodes), with the weighted network diameter - maximum total travel times - of $\approx 54$ minutes, and an average shortest path of about 21 minutes. The average effective velocity is $\approx 28$ km/h and the average waiting time is about $2.7$ minutes. The core of the network exhibits relatively high complexity, with several circuits which are associated with high efficiency and robustness \cite{derrible2010complexity, cats2020metropolitan}. The radial lines then extend to the city periphery.

\subsubsection{Relationship between node centrality in $G$ and $\overline G$}

\begin{figure}[H]
	\begin{center}
		\texttt{}\includegraphics[scale=0.5]{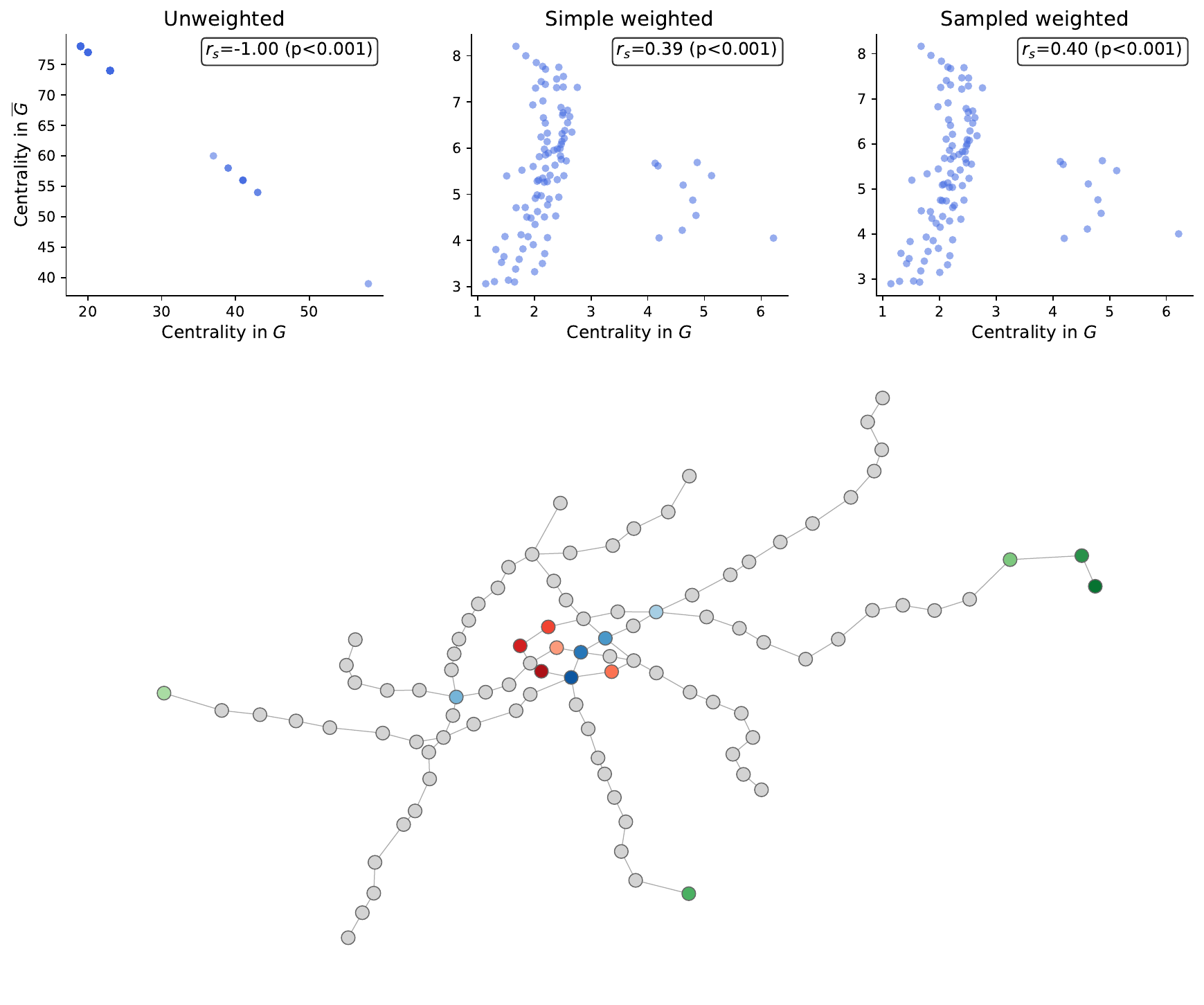}
	\end{center}
	\caption{Vienna: Degree centrality. Top row shows scatter plots between centrality in $G$ ($x$-axis) and $\overline G$ ($y$-axis). From left to right: unweighted, simple weighted, and distance-weighted bootstrapped representations. Each data point represents a stop in the network. The legend in each subplot shows the values of the Spearman correlation coefficient together with $p$-value and distance correlation. Bottom plot shows the L-space representation of the network. The five most important nodes in weighted $G$ are shown in blue (decreasing strength of colour for decreasing centrality) and similar for the five most important nodes in bootstrap weighted $\overline G$ in red. We note that there is no overlap between the two groups, i.e., the most important nodes in $\overline G$ are distinct from the set of the most important nodes in $G$. In addition, the five least important nodes in $G$ are shown in green (darkest green meaning the smallest centrality value). Note that the centralities are calculated in P-space but are visualised in L-space for clarity. }
	\label{fig:ba_cd}
\end{figure}

\begin{figure}[H]
	\begin{center}
		\texttt{}\includegraphics[scale=0.5]{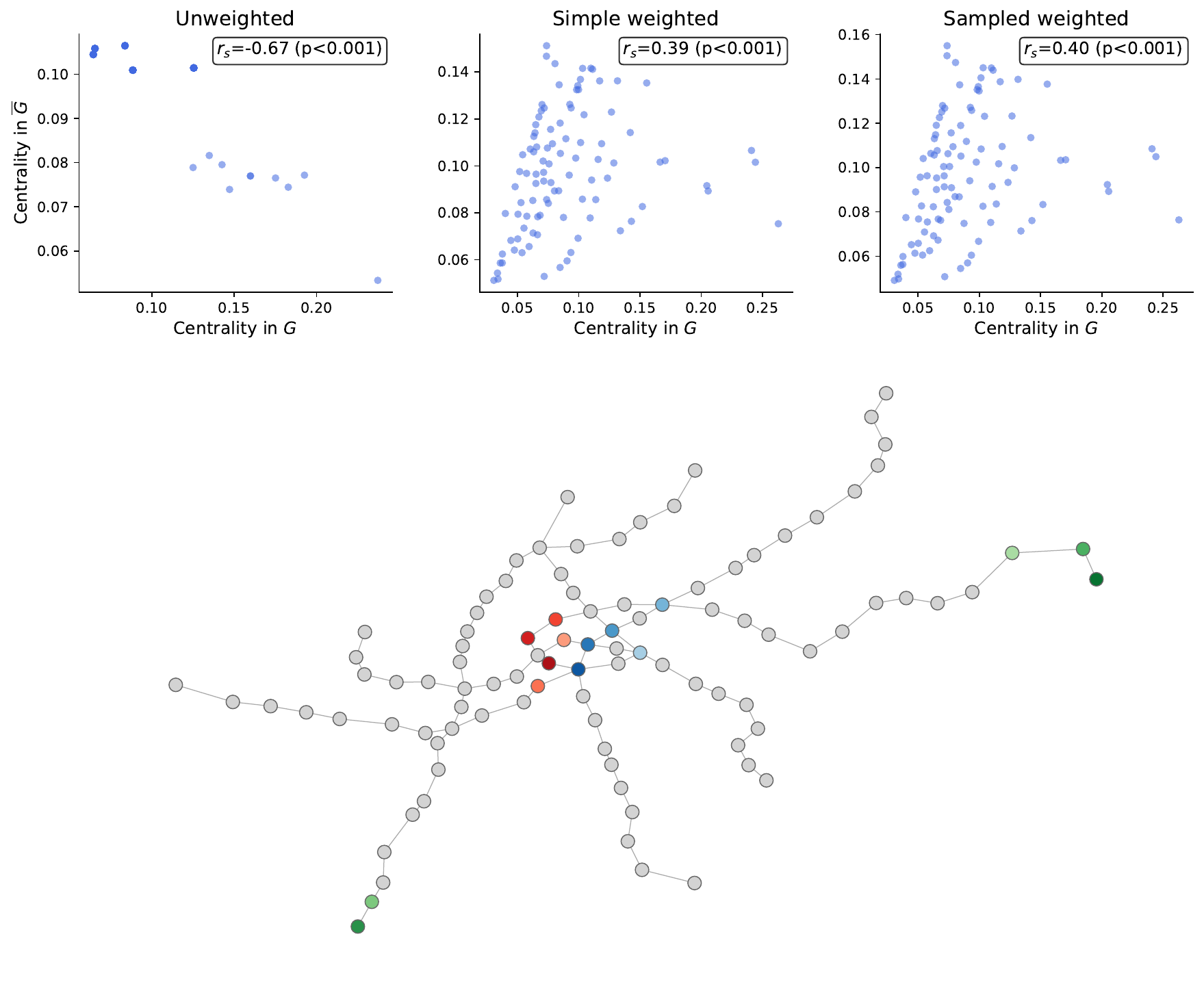}
	\end{center}
	\caption{Vienna: Eigenvector centrality. (See caption in Figure \ref{fig:ba_cd}.)}
	\label{fig:ba_ce}
\end{figure}

The relationships between node centrality in the original graph $G$ and its complement $\overline{G}$ are shown in Figure \ref{fig:ba_cd} for degree centrality and \ref{fig:ba_ce} for eigenvector centrality. 
In each of the figures, the top row shows scatter plots of respective centrality for $G$ ($x$-axis) and $\overline G$ ($y$-axis) for two cases: $(i)$ unweighted $G$ and unweighted $\overline G^u$ (upper left plot), and $(ii)$ weighted $G$ and sampled weighted $\overline G^s$ (upper right plot). In the bottom row, the L-space representation of the network with the five most central nodes in $G$ and $\overline G^s$ highlighted in blue and red colours, respectively (decreasing in darkness of each colour with decreasing centrality). Note that the centralities are calculated in the P-space representation, and the visualisations are in L-space for clarity.

Remarkably, the most central nodes identified in the weighted complement graph are those in, or near, the geographical and topological centre. Most notably, they represent the geographically central nodes that lack in topological connectedness. The nodes that are identified as the most central in $\overline G^s$ are not those that are the least important in $G$. Indeed, the least important nodes, as marked in green colour in Figures \ref{fig:ba_cd} and \ref{fig:ba_ce}, are those on the geographical (and topological) peripheries, as expected. This is further observed in the scatter plots in the upper rows. For an unweighted graph and its complement, the relationship between node degree in both representations will be a negative linear function by definition (i.e., the complement graph includes all the missing edges of $G$), which is indeed seen in the upper left plot of Figure \ref{fig:ba_cd}. For the unweighted eigenvector centrality, the relationship will still be negative although not strictly linear (upper left plot in Figure \ref{fig:ba_ce}). In contrast, for the weighted versions, the relationship between both types of centrality is positive for the majority of the nodes, with a set of outlier nodes, visible especially for degree centrality (upper right plots in Figures \ref{fig:ba_cd} and \ref{fig:ba_ce}). We also note that the results are similar for the simple weighted and bootstrapped averaged distance-based sampled weighted representations, indicating a level of robustness with respect to the specific weighting scheme used.

Outside of this general shape on the left-hand side of the plots (i.e., nodes with lower centralities in $G$), there is a group of outliers in the middle to bottom right-hand side of the plots. The data points in this group correspond to the set of the most central nodes in $G$, and their centrality is in most cases significantly larger than that of the rest of the nodes. The low centrality values of these nodes in $\overline G^s$ indicate that they are very tightly connected to all other nodes in the original network, indicating a level of saturation in connectedness, and therefore have a low potential for new connections. 

We also observe that different, albeit nearby, nodes are identified as the most central according to both centrality measures. Degree centrality directly measures the number of nodes reachable without transfers, weighted by the inverse of total travel time. Eigenvector centrality also incorporates information about the importance of neighbours, providing a global view. Note that three of the five most central nodes in $\overline G^s$ lie on line U2, which services fewer stops than other lines. Similarly, another of the most central nodes in $\overline G^s$ lies on U3 which is also short compared to the longest lines U1 and U4.

Similar results are obtained for all 31 networks in the dataset and the plots similar to those in Figures \ref{fig:ba_cd} and \ref{fig:ba_ce} are included in the supplementary material.




The results for the weighted networks might appear counter-intuitive at first, and we posit that the explanation is provided by the spatial positioning of the nodes. The nodes in $G$ with higher importance are those that are close to both the topological and geographical centre of the network. The nodes that are close to topological intersections or lie on longer routes (i.e., routes with more nodes) have higher centrality values, fortified by their incident edges carrying lower costs compared to the nodes on the outskirts. The same phenomenon occurs in the complement graph, where a high number of edges is incident to the non-transfer nodes with low edge costs due to their geographical positions. This translates to high connection strengths, and consequently to high centrality of these nodes.

We test this hypothesis in the next section, where we compare the correlations between node centrality in $G$ and $\overline G^s$ against several null models that break the topology-geography coupling inherent to the original metro networks.


\subsubsection{Testing against null models}

\begin{figure}[H]
    \centering
    \begin{subfigure}[b]{\textwidth}
        \centering
        \includegraphics[width=1\textwidth]{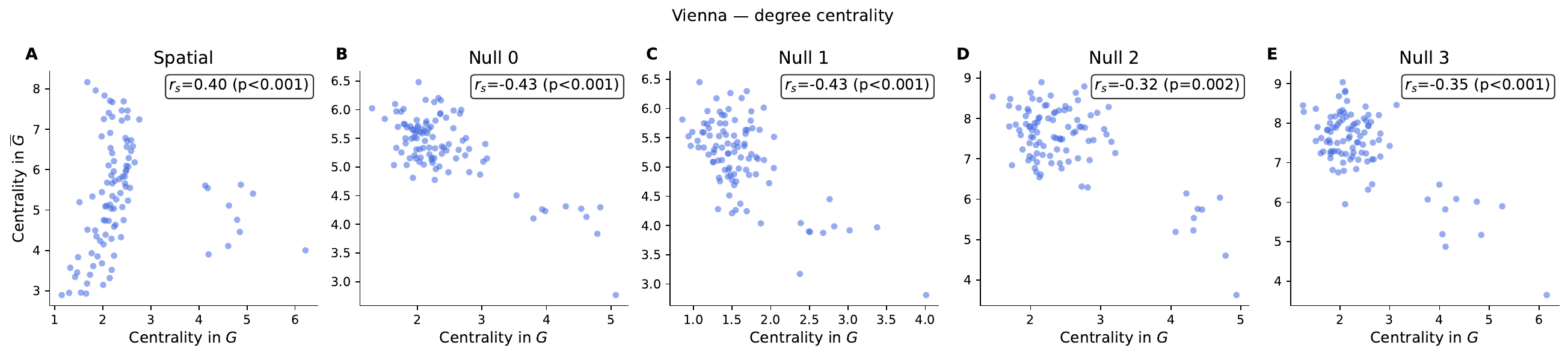}
        \caption{}
        \label{fig:sub1}
    \end{subfigure}
    \vfill
    \begin{subfigure}[b]{\textwidth}
        \centering
        \includegraphics[width=1\textwidth]{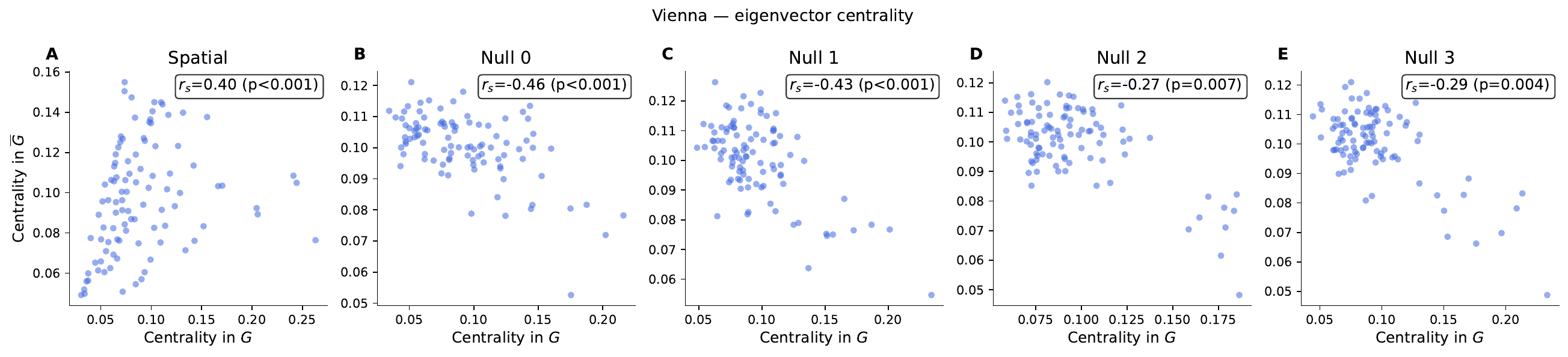}
        \caption{}
        \label{fig:sub2}
    \end{subfigure}
    \caption{Comparison of centrality relationships for the original network and its sampled weighted complement graph (leftmost) to all three null models. Top row: degree centrality; bottom row: eigenvector centrality.}
    \label{fig:null_scatter}
\end{figure}

In this section, we systematically test whether the observed phenomena arise from (exogenously given) spatial features. To this end, we create three null models as described in Methods (Subsection \ref{sec:methods_null}). We then observe the relationships between node centrality in $G$ and $\overline G$ for all null models. The scatter plots for Vienna are shown in Figure \ref{fig:null_scatter}. For the spatial model, we take the original weighted network and its sampled weighted complement graph $\overline G^s$. Note that for these plots, the results are shown for a single case of rewiring. We consider multiple rewirings and confidence intervals for the Spearman correlation coefficient in Section \ref{sec:null_all_cities}, based on which we can assert that the single result here is representative and close to the average values.


We observe that for all null models, the correlation is significant and negative. This is in remarkable contrast to the original network, where the correlations are moderately positive $r_S=0.40$. In the null models, the strongest negative correlations are observed for the Null 0 ($r_S=-0.43$ for degree and $r_S=-0.46$ for eigenvector centrality) and Null 1 models ($r_S=-0.43$), which is expected as these have the most naive weighting schemes. In comparison, Null 2 and Null 3 models are more conservative, with more stringent rules on their construction. We observe that the correlation strengths are somewhat lower compared to Null 1, albeit still significantly negative ($r_S=-0.32$ and $r_S=-0.35$ for degree centrality for Null models 2 and 3, respectively; and $r_S=-0.27$ and $r_S=-0.29$ for eigenvector centrality for Null models 2 and 3, respectively).

\begin{figure}[H]
	\begin{center}
		\texttt{}\includegraphics[scale=0.5]{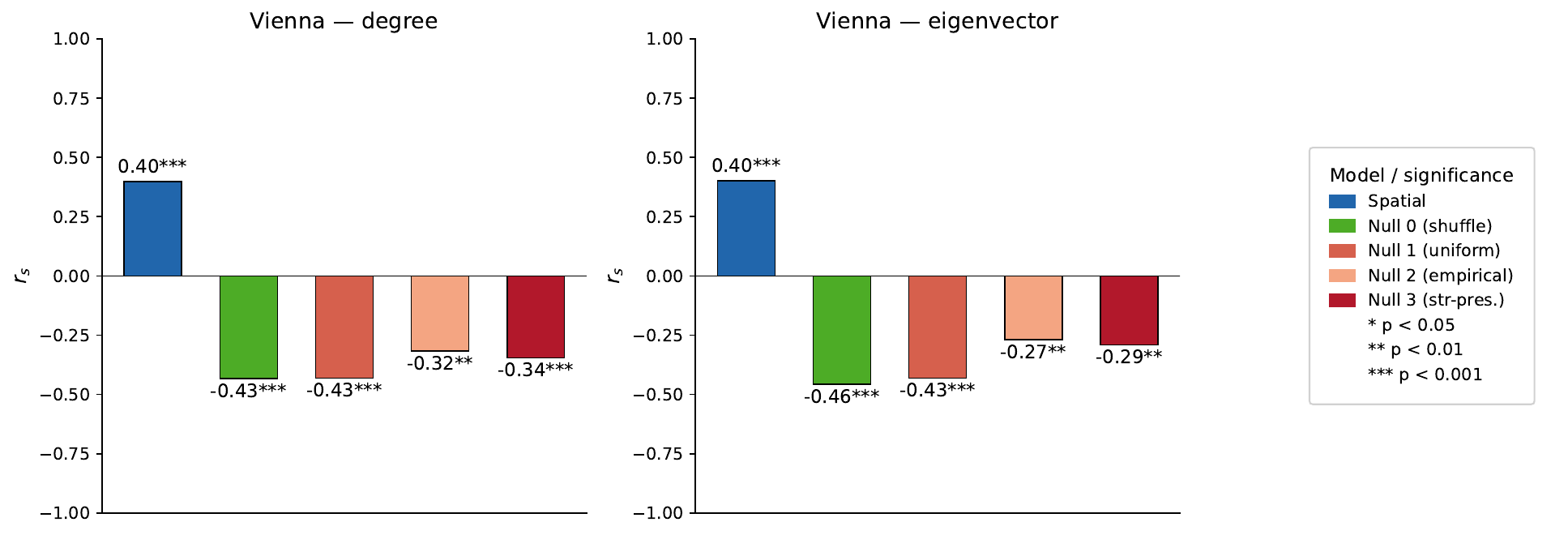}
	\end{center}
	\caption{Null models statistics summary for Vienna.}
	\label{fig:null_summary_ba_mon}
\end{figure}

Focusing here on degree centrality, we note that the correlations of centrality in $G$ and $\overline G$ transition from $(i)$ a strictly negative linear function in the unweighted network (Figure \ref{fig:ba_cd}, top left); through $(ii)$ significantly negative and linear in shape for the Null models, where the weighting is decoupled from space, and finally to $(iii)$ significantly positive in the original network where the weights are determined by spatial distances between nodes. For the eigenvector centrality, a qualitatively similar behaviour of transitions is observed, although even in the unweighted network there is not a clear linear relationship, which carries further to the Null models where the negative correlation strengths are somewhat lower than for degree centrality.
A summary of statistics is shown in Figure \ref{fig:null_summary_ba_mon}.


\subsection{Null model results for all cities}
\label{sec:null_all_cities}

In this section, we test the statistical significance of the differences between the spatial and null models for the full dataset. The data used are the curated datasets of L- and P-space representations of 51 metro networks worldwide \cite{lspace, pspace}. We select a subset of the networks based on the number of nodes: we only take networks with more than 45 nodes to exhibit sufficient complexity of the networks for this analysis, and exclude the largest network (New York City metro with $n=421$ nodes) due to prohibitive computational times. The dimensions of networks included in our analysis range from $n=50$ (e.g. Lisbon, Philadelphia) to $n=303$ (Paris). Their geographical dimensions, as measured by the longest distance between all pairs of nodes, range between 10 and 35 km for most of the networks, with some networks spanning larger distances (e.g., 56 km for Valencia and 75 km for San Francisco). The maximum travel times range from 35 minutes (Lisbon) to about 130 minutes for San Francisco and Valencia. The average effective velocities are between 22 km/h and 35 km/h for most cities. Network density in P-space ranges from 11\% (Paris) to 62\% in Lille and 69\% in San Francisco. 
The key statistics for the 31 networks are given in Appendix in Table \ref{tab:stats_31}. 

The detailed null model analysis as illustrated for Vienna in the previous subsection is provided for all 31 PTNs in the supplementary material and in Appendix \ref{app:summary_null}, Figure \ref{fig:null_summary_all}.

\begin{figure}[H]
    \centering
    
    \begin{subfigure}[b]{0.48\textwidth}
        \centering
        \includegraphics[width=\textwidth]{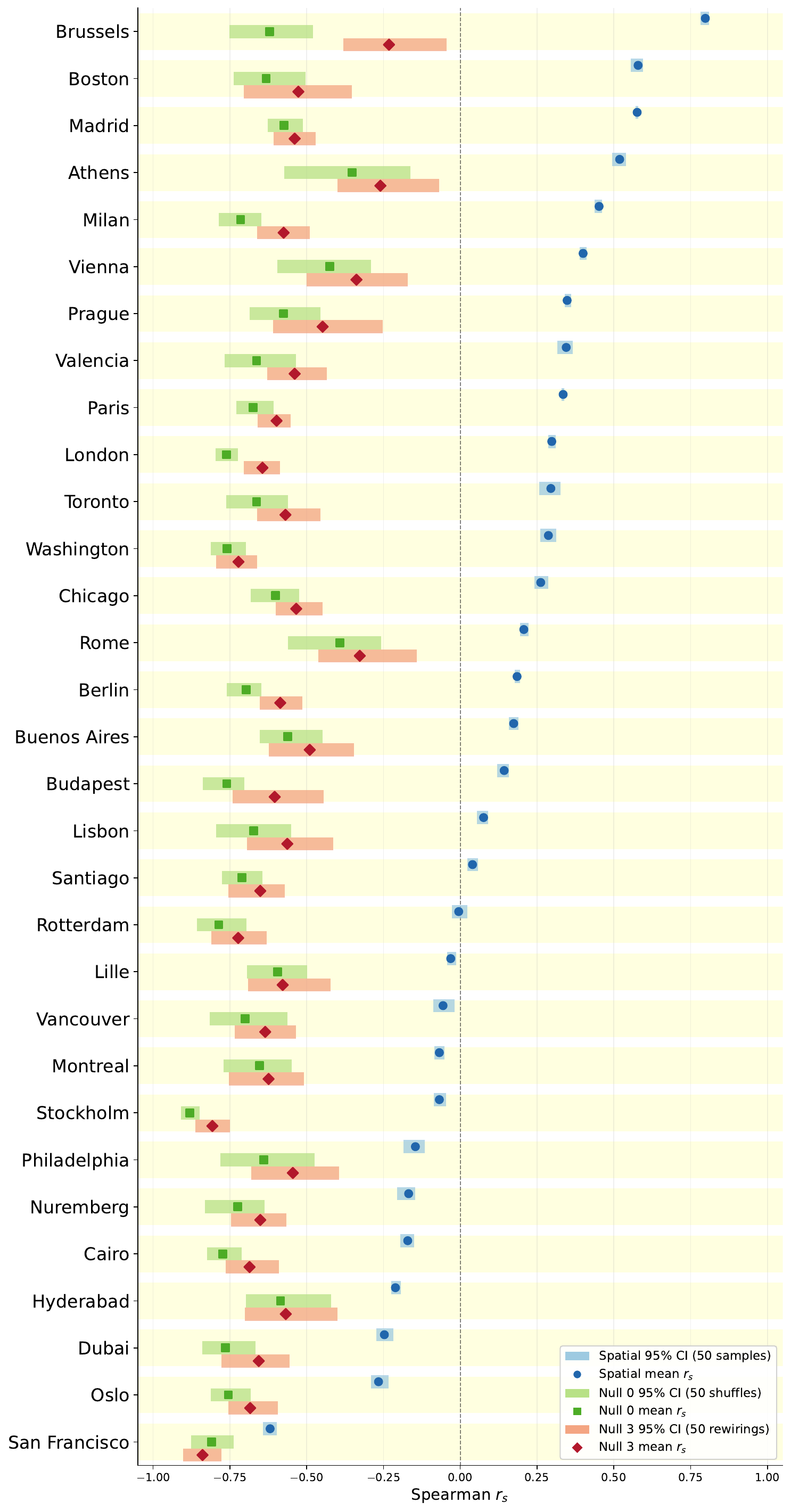}
        \caption{Degree centrality.}
        \label{fig:all_cities_degree}
    \end{subfigure}
    \hfill
    \begin{subfigure}[b]{0.48\textwidth}
        \centering
        \includegraphics[width=\textwidth]{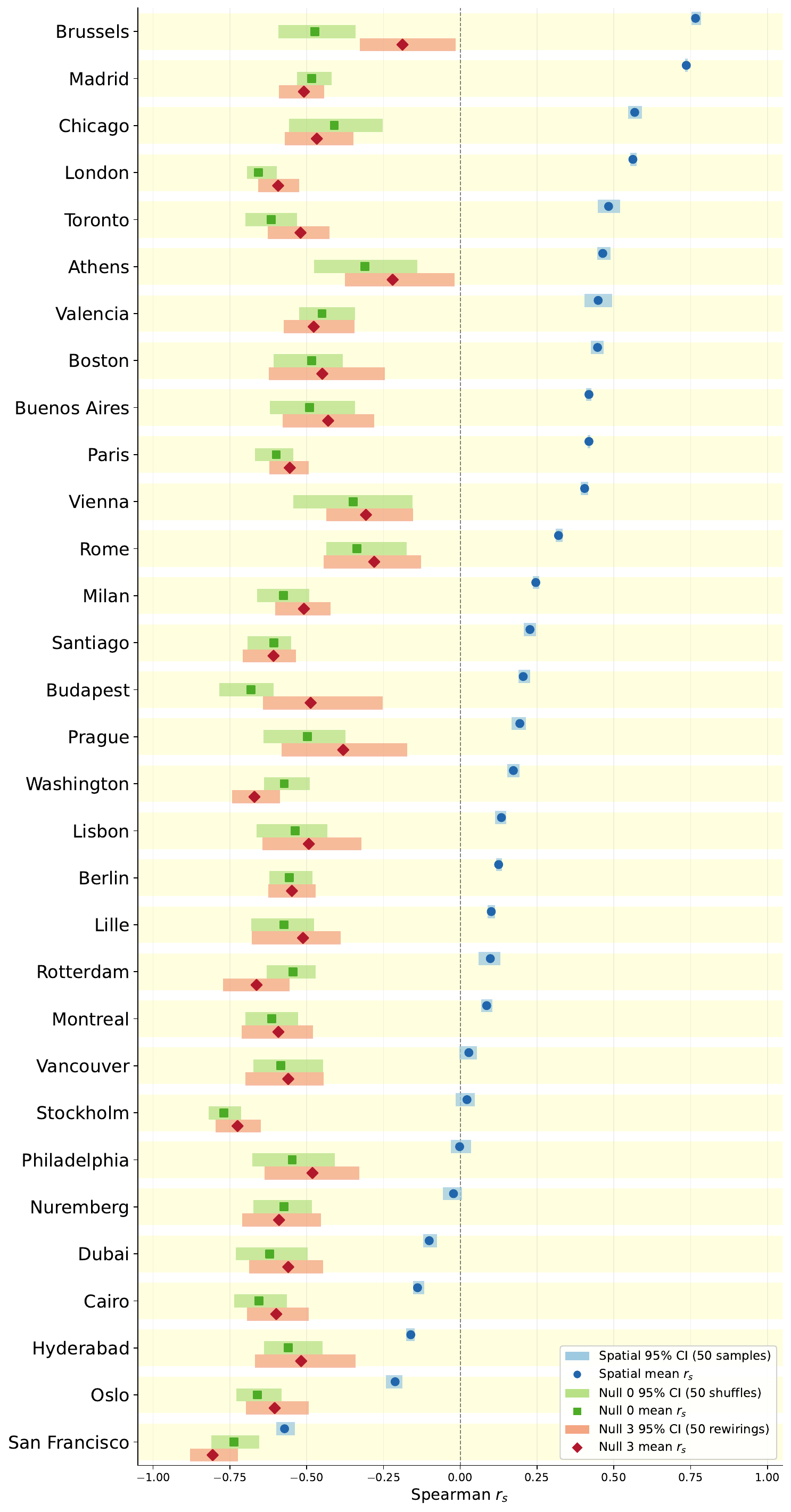}
        \caption{Eigenvector centrality.}
        \label{fig:all_cities_eigenvector}
    \end{subfigure}
    
    \caption{Comparison of Spearman correlation coefficients for node centrality in $G$ and $\overline G$ for (a) degree and (b) eigenvector centrality. The correlation analysis is repeated $N=50$ times for both the spatial model (blue markers) and the Null 0 (green markers) and Null 3 (red markers) models. The markers indicate the average value of $r_S$ over the 50 samples, and are shown with 95\% confidence intervals.}
    \label{fig:all_cities_distributions}
\end{figure}

For the significance testing, we take the $\overline G^s$ spatial model, i.e. with complement edge weights sampled from the network-specific distributions, and perform $N=50$ samplings and calculate the Spearman correlation coefficient $r_S$ for each sampling. We then determine the average value of $r_S$ and its 95\% confidence interval (CI). Similarly, we take the Null 3 model (i.e., the most conservative one) and perform $N=50$ rewirings, then calculate $r_S$ for each rewiring and provide average values with 95\% CI. The results are shown in Figure \ref{fig:all_cities_degree} for degree and Figure \ref{fig:all_cities_eigenvector} for eigenvector centrality.



We see that for all networks, the difference between the spatial and the null models is significant, confirming that the correlation between node importance in $G$ and $\overline G^s$ is caused by spatial effects. The values of $r_S$ for spatial models are positive for a small majority of the networks, with the highest being recorded for Brussels ($r_S\approx 0.8$). For a subset of cities, the correlation is insignificant (e.g., Vancouver, Stockholm, Philadelphia in the case of eigenvector centrality), and for some is is negative (e.g. Cairo, Hyderabad), most significantly so for San Francisco ($r_S\approx -0.6$). We note here that the San Francisco network is an outlier in this dataset with respect to geographical distance as it spans about 75 km, while the values for most of the other networks range between approximately 10 and 30 km. Confidence intervals for $r_S$ in spatial models are narrow for all cities, indicating robustness of the correlation with respect to a single instance of random sampled weighting.

For the Null 0 and Null 3 models, average values and upper CI bounds of $r_S$ are negative for all cities and for both centrality values. The CI widths are wider, indicating large variations across single instances of rewirings. Even for cities with large variations, the CI ranges do not overlap with those of the spatial model, strengthening the case for the coupling between topology and geography being the primary underlying effect for the relationships observed in the spatial models.

\section{Discussion and conclusion}

This study presented a method for constructing weighted complement graphs of spatial networks, and provided a systematic analysis of the relationships between node importance in the original network and its complement. The application of the method to 31 metro networks worldwide revealed highly non-linear relationships between the two, reversing the negative relationship of the unweighted version into a positive one for a majority of the (less important) nodes, with a notable group of outlier nodes, corresponding to the network hubs. While the results may appear counter-intuitive, they reflect the underlying interplay between network topology and geography, which was validated by testing the results against several null models in which the topology-geography relationship is decoupled.

We demonstrate that the proposed method offers a novel way of looking at public transport networks, and, we propose, spatial networks with functional connections in general. Our findings show that the most important nodes in the weighted complement graph are markedly different from the least important nodes in the original network. Instead, they correspond to medium-high importance nodes in the original, indicating that geographically central nodes with low topological connectedness are those with the highest potential for new connections. In the context of transport networks, we interpret this as: the potential for access is not the same as inaccessibility; rather, these are two orthogonal axes along which we can rank nodes or locations.

Our conclusion echo those of \cite{batac2022shortest} who studied the least important nodes in the road network. They found that these are either the nodes at the geographical periphery of the network, or those in the geographical centre that are poorly connected. The latter are also a source of the largest inefficiencies in the network. Our results show a marked similarity to the findings about the latter group of least important nodes in the referenced work, reflecting the importance of spatial scale when considering topological characteristics.

In PTN planning, this carries important implications for line planning and frequency setting problems. The results suggest that planning multiple routes through geographically central nodes has a large potential for increasing overall efficiency. In addition, 
the identification of the nodes with the highest potential for connections may inform the approaches in edge augmentation problems, where the objective is to add $k$ edges to the existing graph to minimise its resistance distance and is known to be NP-hard \cite{predari2022faster, zhou2025efficient}.
The addition of new infrastructural links may serve in distributing flows and therefore congestion effects over the central part of the network, potentially alleviating their ramifications. This also provides a larger number of cycles, which is closely connected to higher network robustness \cite{derrible2010complexity, cats2020metropolitan}. Building a densely connected centre is especially relevant for metro networks as it makes their core - where demand is highest and most transfers take place - robust and accessible for a comparatively small cost (as estimated by the length of new links), while for the peripheral part (radial lines), robustness is more feasibly ensured by offering alternative modes of transport, such as bus lines. Overall, our results are in contrast with the generally accepted view that when planning for more accessible and robust networks, one should focus on the most topologically central nodes in the network.

Alongside the discussed contributions of this study, it is also subject to certain limitations, which we discuss in the following and connect those to prospective venues of future research. First, we note that the analysis was performed only for one type of networks, i.e., metro networks. Beyond PTNs, other spatial networks with functional connections may exhibit high densities, making the complement graph approach useful; e.g., air transport networks, or origin-destination flow networks where the actual flows from data are compared to gravity models. The generality of the findings will need to be tested on a larger set of spatial network classes.

As a methodological limitation, we note that we consider P-space representations, which carries line information, yet in analysing the missing edges, there is no clear correspondence to the structure of potential new lines. This limits the applicability of the method to PTN planning problems. An extension of the analysis may explicitly include line-level information, possibly by using the bipartite B-space representations of stops and lines, or a heterogeneous multilayer network with stops in one layer and lines in the other layer. Such an extension will enable novel insights into the planning problems. Moreover, as hinted above, the analysis of the operational representation does not explicitly consider infrastructure, and the implications for adding new infrastructural links are more akin to educated guessing. Adding an infrastructural layer and connecting it to line planning will offer a more complete picture of the problem.
Furthermore, the method only considers (missing) links amongst the set of existing nodes (stops). An interesting extension will be an investigation of the potential for adding new nodes, potentially based on demand attributes as extracted from land use data.




\section{Methods}

We consider a P-space network, which is formalised as a weighted undirected graph $G$ on $n$ nodes and with $m$ edges. Edge weights $t_{ij}$ represent the total travel time along the edge, and are thus of the cost type. The strength $w_{ij}$ of the connection is then calculated as the inverse of the cost; $w_{ij} = 1/t_{ij}$. We take as its weighted adjacency matrix the matrix of connection strengths; $W = (w_{ij})$.

\subsection{Building the weighted complement network}

We consider three versions of the complement graph; unweighted $\overline G^u$, and two weighted versions: $(i)$ one with simple weighting, using the network-specific average effective velocity and waiting time, $\overline G^a$ (the superscript $a$ denoting the average), and $(ii)$ one with probabilistic weighting where we sample from the network-specific joint distribution of effective speeds and waiting times, conditional on link length, $\overline G^s$ (the superscript $s$ denoting sampling).

\subsubsection{Simple weighting}

For calculating effective speed and waiting times, we need to know stop coordinates, in-vehicle times and service frequencies. This attributes can be extracted from GTFS data that is available for many PTNs worldwide. We do not however know the actual shape and therefore the length of links, so we calculate the effective speeds, which is defined as:
%
\begin{equation}
    v^{eff}_{ij} = \frac{d_{ij}}{t_{ij}},
\end{equation}
Here, $d_{ij}$ is the geographic distance between nodes $i$ and $j$ and $t_{ij}$ is the in-vehicle time between them. The average waiting time $t^{wait}_{ij}$ is calculated as half the headway, or inverse frequency:
\begin{equation}
    t^{wait} = \frac{1}{2f_{ij}},
\end{equation}
where $f_{ij}$ is the service frequency between nodes $i$ and $j$. Note that when there are multiple lines servicing the same pair of stops, the frequencies on those segments are summed over. We then average over all frequencies and effective speeds to obtain the network-specific average frequency $\langle f\rangle$ (from which we obtain the average waiting time $\langle t^{wait}\rangle$) and the average effective speed $\langle{v}^{eff}\rangle$, respectively.

In weighting the complement network, we start from the idealised network, which we define as the complete graph on the given set of nodes, where the missing edges in the original graph - corresponding exactly to the edges of the complement graph - are weighted based on the network-specific values of $\langle v^{eff}\rangle$ and $\langle f\rangle$. We calculate the geographic distance matrix $D$ for the whole network. In this averaged approximation, each complement edge is assigned a travel time weight based on the geographic distance $d_{ij}$ between nodes $i$ and $j$, which is divided by the constant value of the average effective speed, and the constant average waiting time is added:

\begin{equation}
    t_{ij} = \frac{d_{ij}}{\langle v^{eff}\rangle} + \langle t^{wait}\rangle
\end{equation}

\subsubsection{Sampled bootstrapped weighting}

For a more realistic case, we extract the full network-specific empirical parametric joint distributions of random variables $V$ (effective velocity $v^{eff}$) and $W$ (waiting time $t^{wait}$) with respect to link geodesic distance $d$:
\begin{equation}
    f_{V,W}(v^{eff}, t^{wait};\, d)
\end{equation}
The joint distributions are built because, in some of the networks, the correlation between $v^{eff}$ and $t^{wait}$ is non-negligible (positive for a subset of cities and negative for another subset of those cities). For the complement graph weights, we then for each edge randomly sample from the distributions, conditional on edge geographical distance.  We repeat the procedure $N=50$ times and bootstrap the results, taking the average values for each link from $N$ samplings. 

\subsection{Comparison of node importance in $G$ and $\overline{G}$}

In the next step, we calculate the node centrality in the original and complement graphs. Specifically, we observe degree and eigenvector centrality. These are defined as:

\begin{itemize}
    \item \textbf{Degree centrality}: in the weighted case, the total strength of the node, defined as the sum of the strengths of the edges incident to the node: $s_i = \sum_j w_{ij}$.
    \item \textbf{Eigenvector centrality}: defined as $x_i = \frac{1}{\lambda}\sum_jw_{ij}x_j$ where $\lambda$ is the leading eigenvalue of the weighted adjacency matrix $W$ and $x_i$ are the coordinates of its principal eigenvector.
\end{itemize}

We choose degree and eigenvector centrality due to the characteristics of the P-space representation. Degree centrality is a local measure that measures the total strength of connection to the node's direct neighbours. Eigenvector centrality is a global measure that recursively incorporates the importance of neighbouring nodes. P-space representation networks of PTNs usually have a very small diameter (often equals 3 or 4 even in large networks). Moreover, in P-space, most passenger journeys are of length one or two, reflecting journeys of one leg or involving a single transfer, respectively. Centralities based on shortest paths are thus not the first choice for this representation.

We then compare the centrality scores for the original graph and its complement, and calculate the Spearman correlation for both variables.

\subsection{Discerning the spatial effects}
\label{sec:methods_null}

To disentangle topology from geography and assess the impact of geographical scale and distribution on the results, we compare the results to three null models. The basic model is Null 0, in which the topology of both the original and the complement network is retained, and the edge weight are reshuffled randomly so that their correlation with spatial distance is broken. We further test the topology-geography relationship by introducing three additional null models where in addition to randomising edge weights, also the topology is changed. In models Null 1 and Null 2, the original unweighted P-space network is rewired so that the number of links and the degree distribution are preserved \cite{maslov2002specificity}. In model Null 3, the rewiring additionally (approximately) preserves the node strength distribution \cite{ansmann2011constrained}. For each rewiring of the original network, we take its complement graph and assign weights to its edges according to one of weighting strategies described below. 



\begin{enumerate}
    \item \textbf{Null 0}: The topology of $G$ and $\overline G$ is left intact, and the edge weights are randomly reshuffled.
    \item \textbf{Null 1}: The swapped links in $G$ keep the original weights. The weights in the resulting complement graph are sampled uniformly from the range of empirical total travel times.
    \item \textbf{Null 2}: The weights in both the resulting rewired graph and its complement graph are sampled from the distribution of empirical total travel times.
    \item \textbf{Null 3}: The swapped links in $G$ keep the original weights. The weights in the complement graph are sampled from the empirical distributions of total travel times.
\end{enumerate}

Thus, the edge weights in the complement graphs of each rewired network are assigned randomly: either simply reshuffled between existing original edges in Null 0; uniformly from the range of empirical values (Null 1); or sampled from the distribution (Null 2 and 3). The edges in the rewired original network keep the original weights in models Null 1 and Null 3, and are randomly sampled from the distribution in Null 2.

The three null models are designed so that each of them breaks different structural regularities. In Null 1, the topology of $G$ is separated from its geography by changing the structure and keeping the weights. In Null 2, both the topology and weight structure is disrupted, and weights are sampled randomly from the empirical distribution. Null 3 is the most stringent, or conservative, model since it requires, next to preserving the overall degree distribution, to preserve the node strength distribution within 5\% for each node.

\section*{Author contributions}

Conceptualisation, Methodology, Investigation: T.\v S. and O.C.; Software, Formal analysis and Writing - Original Draft: T.\v S.; Writing - Review and Editing: O.C.

\appendix

\section{Summary of key topological and geographical statistics for all 31 networks}

\begin{table}[H]
\caption{Key statistics of the metro networks used in the analysis.}
\label{tab:stats_31}
\begin{tabular}{lrrrrrrrrrr}
\toprule
City & $n$ & $m(G)$ & $d(G)$ & $d(\overline G)$ & $\langle v^{eff}\rangle$ & $\langle t^{wait}\rangle$ & $\langle l\rangle$ &$l^{max}$& $\langle t\rangle$ & $t^{max}$ \\
\midrule
Athens & 61 & 675 & 0.37 & 0.63 & 28.33 & 4.71 & 6.80 & 25.22 & 28.34 & 97.91 \\
Berlin & 174 & 2711 & 0.18 & 0.82 & 26.36 & 13.00 & 6.58 & 23.97 & 28.58 & 86.03 \\
Boston & 52 & 467 & 0.35 & 0.65 & 25.03 & 5.31 & 6.28 & 23.90 & 29.06 & 64.80 \\
Brussels & 59 & 900 & 0.53 & 0.47 & 23.07 & 2.98 & 4.60 & 14.16 & 17.54 & 45.31 \\
Budapest & 48 & 345 & 0.31 & 0.69 & 25.95 & 1.98 & 3.83 & 11.74 & 15.41 & 34.48 \\
Buenos Aires & 78 & 662 & 0.22 & 0.78 & 23.08 & 2.42 & 3.68 & 11.17 & 19.61 & 45.79 \\
Cairo & 61 & 820 & 0.45 & 0.55 & 28.49 & 1.32 & 10.77 & 35.00 & 28.80 & 77.92 \\
Chicago & 137 & 2233 & 0.24 & 0.76 & 27.11 & 6.04 & 8.98 & 33.23 & 40.72 & 97.75 \\
Dubai & 53 & 664 & 0.48 & 0.52 & 34.53 & 2.96 & 12.50 & 38.59 & 33.11 & 83.75 \\
Hyderabad & 56 & 640 & 0.42 & 0.58 & 33.52 & 2.87 & 8.02 & 24.71 & 23.65 & 51.03 \\
Lille & 60 & 1098 & 0.62 & 0.38 & 25.06 & 1.49 & 6.85 & 17.77 & 20.63 & 56.40 \\
Lisbon & 50 & 375 & 0.31 & 0.69 & 31.71 & 1.92 & 3.40 & 9.83 & 13.95 & 34.78 \\
London & 261 & 6439 & 0.19 & 0.81 & 26.97 & 3.81 & 11.93 & 42.65 & 36.50 & 117.54 \\
Madrid & 240 & 3739 & 0.13 & 0.87 & 22.76 & 3.14 & 6.47 & 31.07 & 38.27 & 117.78 \\
Milan & 106 & 1569 & 0.28 & 0.72 & 24.87 & 2.44 & 6.55 & 27.75 & 23.96 & 70.63 \\
Montreal & 67 & 872 & 0.39 & 0.61 & 28.95 & 3.40 & 5.84 & 17.51 & 21.48 & 48.01 \\
Nuremberg & 49 & 555 & 0.47 & 0.53 & 26.39 & 7.36 & 4.84 & 16.09 & 17.28 & 40.57 \\
Oslo & 101 & 2176 & 0.43 & 0.57 & 22.67 & 13.42 & 6.64 & 19.73 & 34.71 & 74.27 \\
Paris & 303 & 5206 & 0.11 & 0.89 & 20.89 & 1.94 & 4.61 & 15.88 & 21.53 & 60.38 \\
Philadelphia & 50 & 634 & 0.52 & 0.48 & 29.71 & 6.09 & 5.99 & 16.80 & 24.68 & 54.38 \\
Prague & 58 & 602 & 0.36 & 0.64 & 38.30 & 2.08 & 6.34 & 21.31 & 16.80 & 38.40 \\
Rome & 73 & 879 & 0.33 & 0.67 & 26.13 & 3.14 & 5.84 & 16.55 & 28.38 & 72.34 \\
Rotterdam & 70 & 1081 & 0.45 & 0.55 & 31.75 & 5.66 & 9.65 & 30.98 & 33.49 & 86.74 \\
San Francisco & 50 & 844 & 0.69 & 0.31 & 54.29 & 10.69 & 27.12 & 75.94 & 47.78 & 127.76 \\
Santiago & 119 & 1481 & 0.21 & 0.79 & 37.69 & 3.18 & 6.79 & 21.41 & 23.19 & 60.73 \\
Stockholm & 101 & 1621 & 0.32 & 0.68 & 25.78 & 9.78 & 7.57 & 20.73 & 30.85 & 76.31 \\
Toronto & 75 & 1190 & 0.43 & 0.57 & 25.43 & 2.67 & 7.34 & 24.31 & 29.12 & 79.28 \\
Valencia & 95 & 1829 & 0.41 & 0.59 & 30.42 & 8.36 & 12.71 & 56.55 & 39.58 & 126.88 \\
Vancouver & 52 & 503 & 0.38 & 0.62 & 35.27 & 3.24 & 8.58 & 23.00 & 30.23 & 75.42 \\
Vienna & 98 & 1138 & 0.24 & 0.76 & 28.14 & 2.67 & 4.95 & 15.55 & 20.93 & 53.50 \\
Washington & 89 & 1368 & 0.35 & 0.65 & 31.07 & 6.31 & 10.64 & 43.20 & 37.58 & 85.61 \\
\bottomrule
\end{tabular}
\caption*{The variables are: $n$ - number of nodes; $m(G)$ - number of edges in $G$; $d(G)$ - edge density in $G$; $d(\overline G)$ - edge density in $\overline G$; $\langle v^{eff}\rangle$ - average effective velocity; $\langle t^{wait}\rangle$ - average waiting time; $\langle l\rangle$ - average geodesic distance; $l^{max}$ - maximum geodesic distance; $\langle t\rangle$ - average total travel time; $t^{max}$ - maximum total travel time.}
\end{table}

\section{Summary of null model testing for all cities}
\label{app:summary_null}

\begin{figure}[H]
	\begin{center}
		\texttt{}\includegraphics[scale=0.275]{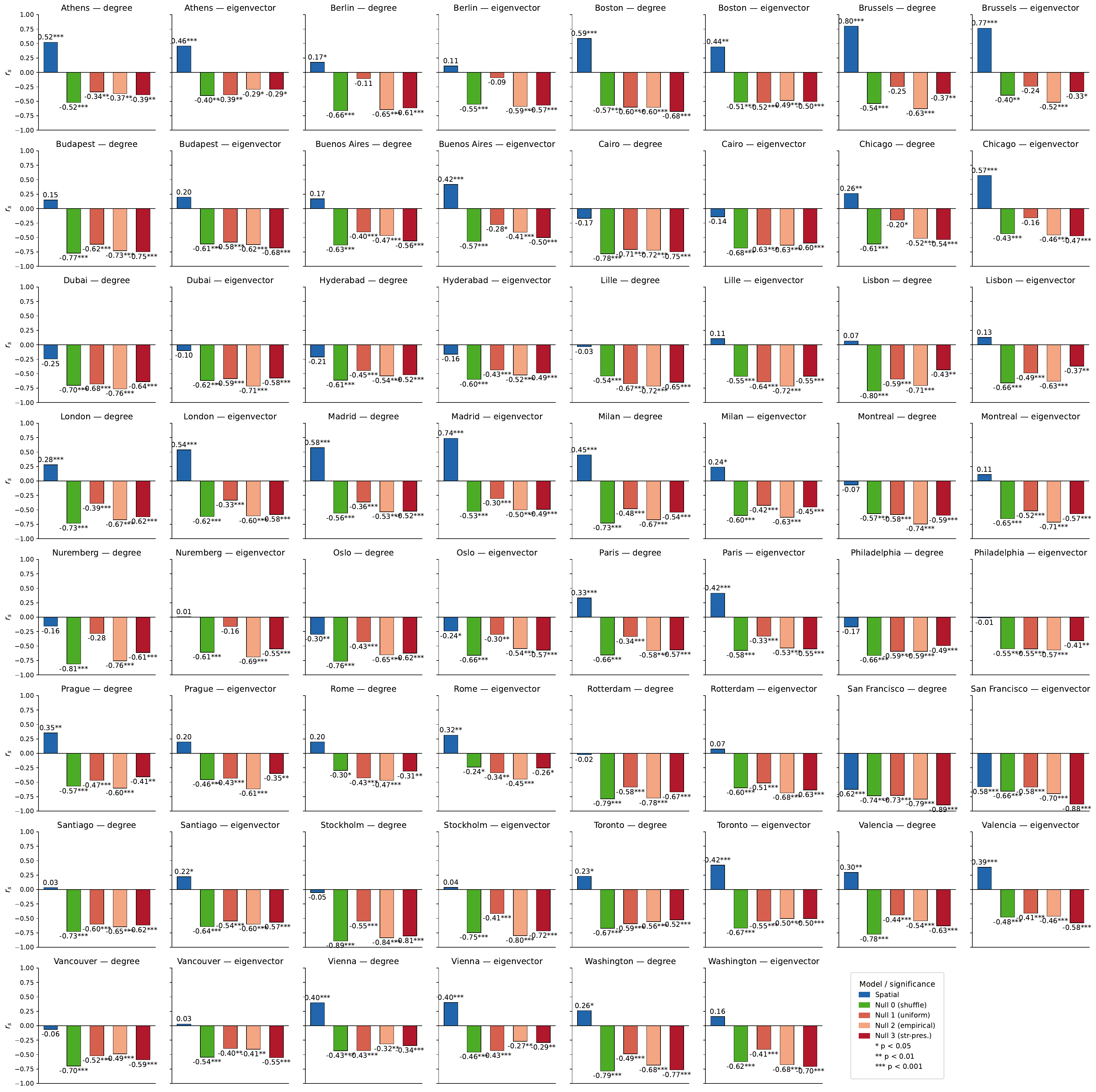}
	\end{center}
	\caption{Null models summary.}
	\label{fig:null_summary_all}
\end{figure}

\bibliographystyle{ieeetr} 
\bibliography{bibliography.bib}

\end{document}